\documentclass[useAMS,usenatbib]{mn2e}

\title{Exact solutions for the populations of the n-level ion}
\author[Michael Taylor and Jos\'e Manuel V\'ilchez]
{Michael Taylor$^{1}$\thanks{michael@damir.iem.csic.es}
 and Jos\'e Manuel V\'ilchez$^{2}$\thanks{jvm@iaa.es}\\
$^{1}$ Departamento de Astrof\'isica Molecular e Infrarroja(DAMIR), Instituto de Estructura de la Mat\'eria(IEM), CSIC, Madrid, Spain\\
$^{2}$ Departamento de Astrof\'isica Extragalactica, Instituto de Astrof\'isica de Andalucia (IAA), CSIC, Granada, Spain.\\}

\begin{document}

\date{Accepted XXXX Month XX. Received XXXX Month XX; in original form XXXX Month XX}

\pagerange{\pageref{firstpage}--\pageref{lastpage}} \pubyear{2006}

\maketitle

\label{firstpage}

\begin{abstract}
We present a matrix solution to the full equations of statistical equilibrium that give the energy level populations of collisionally-excited ions in photoionised gaseous nebulae. The rationale for such a calculation is to maintain a parity between improvements in the quantum-mechanically evaluated values for collision strengths and transition probabilities from the Iron and Opacity Projects on the one hand, and 3D photoionisation codes such as MOCASSIN and astrophysical software for producing nebular diagnostics such as the \textit{Nebular} package for IRAF, on the other. We have taken advantage of the fact that mathematics programs such as MATLAB and Mathematica have proven to be very adept at symbolic manipulation - providing a route to exact solutions for the n-level ion. In particular, we have avoided the substitution of estimated values. We provide the matrices for the 5-level ion as an example and show how the equations faithfully reduce to the exact solution for the 3-level ion. Through the forbidden line ratio $R_{23}$, we compare the exact solution with a) that obtained from the observed emission of the spherical planetary nebula Abell 39, b) 3D Monte-Carlo photoionisation modelling of the same nebula, c) the approximate 5-level program TEMDEN and d) the exact 3-level ion.  The general solution presented here means that programs for the calculation of level populations can obtain solutions for ions with a user-specified number of excited levels. The use of a separate and updatable database of atomic and ionic constants such as that provided by NIST, means that software of more general application can now be made available; particularly for the study of high excitation objects such as active galactic nebulae (AGNs) and supernovae (SNs) where higher excited levels become significant.

\end{abstract}


\section{Introduction}
Essentially 3 parameters fully determine the physical nature of an ionised nebula: the electron temperature ($T_{e}$), the electron density ($n_{e}$) and the level populations of the ions ($N_{i}$). From the latter, ratios of emission rates and emission line intensities can be calculated. The 3 parameter family forms a closed set whereby knowledge of any two of them allows for a determination of the third. In principle, all other secondary physical quantities such as ionisation parameters, ionic abundances and effective temperatures can then be calculated. Until now the exact functional relation between the parameters has evaded astronomers. In this paper, we present a solution to this problem. The context of great progress in observational astrophysics has provided empirical tools to estimate the values of $n_{e}$ (Menzel, Aller \& Hebb 1941) and $T_{e}$ (Aller, Ufford \& Van Vleck 1949; Seaton 1954) and various indicators of ionic abundances (P\'erez-Montero \& D\'iaz 2005). Observations provide access to the properties of the ionised gas through their emission line spectra. In particular, certain ratios of forbidden lines emmited by the $p^{2}$ and $p^{3}$ configuration ions present in the nebular gas, allow the integrated electron density and electron temperature (respectively) to be estimated point to point across projected images. Furthermore, sums of forbidden lines emitted by different stages of ionisation of the gas (abundance indicators) have been found to weakly correlate with metallicity. However, the underlying reason for such correlations has yet to be totally resolved. In this paper we show that, since the emission line intensities depend on the level populations of the ions which in turn depend on the local electron density and temperature, then such correlations  simply reflect an empirical relation between $T_{e}$, $n_{e}$ and $N_{i}$ of the form,

\begin{equation}
	f\left(n_{e},T_{e},N_{i}\right)=0.
\end{equation} 
 
\noindent
Using linear algebra, we have re-cast the equations of statistical equilibrium for the level populations of the ions in the form of a well-posed matrix equation that we subsequently solved using symbolic manipulation to obtain the exact form of the function $f$.\\
\newline
The early work in this field was performed by Menzel(1941, 1962), Aller(1949), Seaton(1954, 1975) and Osterbrock(1967) who obtained initial estimates of the level populations of 3-level ions. The full solution for the 3-level ion was finally worked out by Seaton(1975) and a first order approximation the 5-level ion is currently incorporated into astrophysical software such as TEMDEN (De Robertis et al. 1987). We use these results as well as detailed spectrophotometric observations and 3D Monte Carlo photoionisation modelling of the spherical planetary nebula Abell 39 to validate and compare with the exact solution we have obtained. The theoretical context for the physics of collisionally-excited ions is the thermal equilibrium present in ionised gases. In the next section we highlight the role played by collisionally-excited emission lines (and therefore of the n-level ion) to the thermal balance of ionised gaseous nebulae.

\section{Thermal Equilibrium}
The temperature in a static nebulae is fixed by the equilibrium between heating by photoionisation and cooling by recombination, free-free radiation (bremsstrahlung) and line radiation. In ionisation equilibrium, photoionisations are balanced by an equal number of recombinations of thermal electrons. The difference between the mean energy of a newly created photoelectron and the recombining thermal electron represents the net gain in energy of the electron gas per ionisation process. In equilibrium, this net gain is balanced by the energy lost by bremsstrahlung and radiation produced by collisional excitation of bound levels of abundant ions that subsequently emit photons leaving the nebula. For radiatively-ionised, low density plasmas, the relaxation times of relevant physical processes are such that we can usually regard a given volume element as being in a steady state. If we can neglect mechanical effects, particularly energy lost or gained by expansion and compression, we can impose the condition of radiative equilibrium (or energy balance) such that the energy absorbed in each volume element then equals the amount emitted (Spitzer 1948; Osterbrock \& Ferland 2006).\\
\newline
Menzel and co-workers first set up the equation of energy conservation (Menzel et al. 1941) whereby the energy absorbed by photoionisation of the gas ($G$) is balanced by the energy liberated in capture and subsequent recombination events ($L_{R}$), free-free Bremsstrahlung emission ($L_{ff}$) and energy emitted in collisionally-excited radiative cooling ($L_{C}$),
\begin{equation}
	G=L_{R}+L_{ff}+L_{C}.
\end{equation}
Closed forms for $G$, $L_{R}$ and $L_{ff}$ can be found in standard texts on the physics of ionised gaseous nebula (e.g. Aller 1984; Osterbrock \& Ferland 2006). What interests us here in the present work is the form of $L_{C}$, the cooling due to the emission of collisionally-excited lines (CELs) and their predominance over the estimation of ionic tempertures. At densities higher than the critical density $N_{c}$,
\begin{equation}
	N_{c}(i)=\sum_{j<i}A_{ij}/\sum_{j\neq i}q_{ij},
\end{equation}
collisional de-excitation becomes important and the cooling rate at a given temperature is decreased. Hence, for $n_{e}<N_{c}(i)$, collisional de-excitation of level $i$ is negligible - the thermal regime of the vast majority of planetary nebulae.\\

\section{Energy loss by CELs}
In many gaseous nebulae, mechanical or magnetic energy from hydromagnetic waves is dissipated in the gas and there can exist regions where the mean temperature is raised such that atomic levels are excited through collisions with electrons. Furthermore, in the presence of sources of high energy UV photons such as those produced in stellar atmospheres or in the radiation fields of stellar clouds, thermalised electrons are capable of ionising the low-lying levels of the ions of heavy elements. As these ions return to their ground states, they emit \textit{permitted} line photons (that fall in the UV domain observable with satellite telescopes). However, atomic term structure, elemental abundances and ionic concentrations are such that in gaseous nebulae, very few collisionally excited permitted lines are observed in optical spectral regions, particularly in the low density plasmas of many nebulae and HII regions. The most frequently observed optical lines come from ions excited by impacts with electrons - the \textit{forbidden} lines that violate the Laporte Parity Rule (Aller 1984). For these lines, we need to know the transition probabilities ($A_{ij}$) and the collision strengths ($\Omega_{ij}$) for the excitation of metastable levels. Forbidden lines cannot be observed in the laboratory and so we have to rely entirely on theory to determine $A_{ij}$ and $\Omega_{ij}$. Observational checks on certain line intensity ratios are sometimes possible, but for most of the transitions on which we depend for nebular plasma diagnostics, observational or experimental checks are not available. In the next section therefore, we rigorously solve for the populations of the n-level ion that give rise to emitted forbidden lines.\\

\subsection{The n-level ion}
Assuming that level populations $N_{i}$ are determined by spontaneous emission (with transition probability $A_{ij}$ per second), i.e. we note that the only possible radiative transitions are downward ones resulting from impacts with thermal electrons, then the formal general equations of statisitcal equilibrium for an ion with excited levels ($i$) above ground having excitation rate coefficients ($q_{ji}$) and de-excitation rate coefficients ($q_{ij}$) to the other levels $j\neq i$ are given by Aller(1984) equation 5-32), 
\begin{eqnarray}
	\sum_{j\neq i}N_{j}n_{e}q_{ji}+\sum_{j>i}N_{j}A_{ji}=\sum_{j\neq i}N_{i}n_{e}q_{ij}+\sum_{j<i}N_{i}A_{ij}
\end{eqnarray}
subject to the condition that the sum over all level populations (i.e. over all stages of ionisation) is equal to the total number density ($N$) of the ions [$cm^{-3}$],
\begin{equation}
	\sum_{j}N_{j}=N.
\end{equation}
Here we have adopted the excitation potential convention $E_{i}>E_{j}$ such that the transition $i\rightarrow j$ corresponds to de-excitation. The balance is between terms that contribute to the population of an ionised level and those that subtract from it. The first term on the left hand side therefore represents collisional de(excitations) to $i$ from other levels ($j\neq i$) while the second term includes radiative transitions to $i$ from upper levels ($j>i$). The left hand side therefore depends on the populations of levels other than $i$. The first term on the right hand side represents collisional de-excitation of other levels by electrons from $i$ while the second term represents downward radiative transitions from level $i$ to lower levels ($j<i$). The right hand side therefore depends on the population of level $i$.\\
\newline\noindent
The next step is to note that $q_{ij}$ is related to the collisional excitation rate $c_{ij}$ [$s^{-1}$],
\begin{equation}
	c_{ij}=n_{e}q_{ij}\equiv Kx\frac{\Omega_{ji}}{\omega_{i}}
\end{equation}
with $K=8.629\times10^{-6}$, Seaton variable $x=n_{e}/T_{e}^{1/2}$ and where $\Omega_{ji}$ is the velocity-averaged collision strength (Seaton 1968),
\begin{equation}
	\Omega_{ji}=\int_{0}^{\infty}\Omega(ji;E)e^{-E/k_{B}T_{e}}d\left(\frac{E}{k_{B}T_{e}}\right),
\end{equation}
for colliding electrons having inital kinetic energy $E=mu^{2}/2$. Here, $T_{e}$ is the electron temperature [$K$] and $n_{e}$ is the electron density [$cm^{-3}$]. The collision strengths must be calculated quantum-mechanically and consist, in general, of a part that varies slowly with energy superimposed with resonance contributions that vary rapidly. Early attempts at calculating them using the Born-Oppenheimer approximation turned out to violate the Mott-Bohr-Peierls-Placzek conservation theorem (Aller 1984), and it wasn't until the late 1950s and early 1960s that Seaton successfully calculated reliable cross-sections using the "Exact Resonance Method" and the "Distorted Wave Method" (Seaton 1968, 1975). With the development of the close-coupling approximation, consistency arguments and, most importantly, the fact that integration over a broad Maxwellian distribution of electron energies tends to dramatically smooth out variations, $\Omega_{ji}$ is now known to be fairly insensitive to temperature (Osterbrock \& Ferland 2006).\\
\newline\noindent
The collisional de-excitation rate,
\begin{equation}
	c_{ij}=n_{e}q_{ji},
\end{equation}
is related to the collisional excitation rate through,
\begin{equation}
	c_{ji}=c_{ij}\frac{\omega_{i}}{\omega_{j}}\epsilon_{ij},
\end{equation}
with $\epsilon_{ij}=e^{-\left(E_{i}-E_{j}\right)/k_{B}T_{e}}$ and where $E_{i}$ and $E_{j}$ are the excitation potentials of the levels. We see that, through the collisional de(excitation) rates and their dependence on the Seaton variable $x(n_{e},T_{e})$ and $\epsilon_{ij}(T_{e})$, the level populations (and therefore all quantities derived from them such as emission line ratios) have the dependency on both electron density and electron temperature given in equation 1. In contrast, the radiative transition probabilities $A_{i,j}$ are independent constants being inversely proportional to the occupancy lifetimes of the upper level. We note in passing that for many nebulae the Seaton variable $\approx10^{2}$.\\
\newline\noindent
Replacing all $n_{e}q_{ji}$ and $n_{e}q_{ij}$ terms in the equations of thermal equilibrium for the excited levels by the expressions above for the collisional excitation and de-excitation rates and dividing through by $Kx\neq 0$ we obtain,
\begin{eqnarray}
\sum_{j\neq i}N_{j}\frac{\Omega_{ji}\epsilon_{ij}}{\omega_{j}}+\sum_{j>i}N_{j}\frac{A_{ji}}{Kx}-\sum_{j\neq i}N_{i}\frac{\Omega_{ji}}{\omega_{i}}-
\sum_{j<i}N_{i}\frac{A_{ij}}{Kx}=0.
\end{eqnarray}
We therefore have a set of simultaneous equations in $n$ unknowns for the ($n-1$) excited levels above ground ($i=1$), supplemented by the total ion density condition $N_{1}+N_{2}+\cdots+N_{n}=N$. Taken together as a set, they have the mathematical form,
\begin{equation}
    \left [ \begin{array}{ccccc}
    1          &1          &1                   &\cdots&1\\
    \alpha_{21}&\alpha_{22}&\alpha_{23}&\cdots&\alpha_{2n}\\
    \alpha_{31}&\alpha_{32}&\alpha_{33}&\cdots&\alpha_{3n}\\
    \alpha_{41}&\alpha_{42}&\alpha_{43}&\cdots&\alpha_{4n}\\
    \alpha_{51}&\alpha_{52}&\alpha_{53}&\cdots&\alpha_{5n}\\
    \vdots     &\vdots     &\vdots       &\vdots\\
    \alpha_{n1}&\alpha_{n2}&\alpha_{n3}&\cdots&\alpha_{nn}\\
\end{array} \right ] \left [ \begin{array}{c} 
    N_{1}\\
    N_{2}\\
    N_{3}\\
    N_{4}\\
    N_{5}\\
    \vdots\\
    N_{n}\\
\end{array} \right ] = \left [ \begin{array}{c} 
    N\\
    0\\
    0\\
    0\\
    0\\
    \vdots\\
    0\\
\end{array} \right ].
\end{equation}
where the $\alpha_{ij}$ are the coefficients of the $N_{i}$. The first row with $\alpha_{1j}=1$ reflects the total ion density condition (equation 5). Introducing the matrix $\tilde{A}$ of coefficients $\alpha_{ij}$, the vector $\textbf{y}$ of level populations $N_{i}$ and the vector $\textbf{b}$ for the right hand side, we can write the simultaneous equations as the linear matrix system, 
\begin{equation}
	\tilde{A}\textbf{y}=\textbf{b}.
\end{equation}
Provided that $\left|\tilde{A}\right|\neq 0$ then we can solve for $\textbf{y}$, 
\begin{equation}
	\textbf{y}=\tilde{A}^{-1}\textbf{b}
\end{equation}
using symbolic manipulation software such as MATLAB or Mathematica. The sought-after level populations are identically given by,
\begin{eqnarray}
	N_{1}&=&y_{1} = (\tilde{A}^{-1})_{11}N\nonumber\\
	N_{2}&=&y_{2} = (\tilde{A}^{-1})_{21}N\nonumber\\
	N_{3}&=&y_{3} = (\tilde{A}^{-1})_{31}N\nonumber\\
	N_{4}&=&y_{4} = (\tilde{A}^{-1})_{41}N\nonumber\\
	N_{5}&=&y_{5} = (\tilde{A}^{-1})_{51}N\nonumber\\
  &\vdots&\nonumber\\
	N_{n}&=&y_{n} = (\tilde{A}^{-1})_{n1}N.
\end{eqnarray}
Finally, the collisionally-excited radiative cooling rate ($L_{C}$), central to the overall thermal balance of gaseous nebulae, can then be calculated from the level populations and is given by,
\begin{equation}
	L_{C}=\sum_{i}N_{i}\sum_{j<i}A_{ij}h\nu_{ij},
\end{equation}
where $h\nu_{ij}$ is the difference in energy level potentials.\\
\newline\noindent
Equations 10-14 allow the astrophysicist to model (exactly) an ion of as many levels as required, provided $\left|\tilde{A}\right|\neq 0$ such that the matrix inverse $\tilde{A}^{-1}$ exists. In the next section we make this more explicit by providing the matrix equations for the 5-level ion.

\subsection{The 5-level ion}
Ions having $p^{2}$, $p^{3}$ and $p^{4}$ electron configurations all have 5 low-lying energy levels. For such ions, collisional and radiative transitions can occur between any of the levels and excitation and de-excitation cross-sections as well as collision strengths exist between all pairs of levels. A central assumption that has historically been made up until now is that only these 5 levels are physically relevant to a  calculation of the observed emission lines of the ion. The justification (Osterbrock \& Ferland 2006) is that higher levels in these ions are not significantly populated through collisions, recombinations or other mechanisms. However, for strongly ionised nebulae in AGNs or SNs for example, the higher levels may indeed be significant to the emission line spectrum.\\
\newline
For a 5-level atom in a steady state with $E_{5}>E_{4}>E_{3}>E_{2}>E_{1}$, the total number density condition for each ion species is given by,
\begin{equation}
	N_{1}+N_{2}+N_{3}+N_{4}+N_{5}=N.
\end{equation}
For the four excited levels above ground, the equations of statistical equilibrium (equation 10) with statistical weights $\omega_{k}=(2k+1)$ give rise to the following four exact level population equations:
\begin{eqnarray}
		N_{1}\frac{\Omega_{21}\epsilon_{21}}{3}\nonumber\\
		-N_{2}\left(\frac{A_{21}}{Kx}+\frac{\Omega_{12}+\Omega_{32}+\Omega_{42}+\Omega_{52}}{5}\right)\nonumber\\
	+N_{3}\left(\frac{A_{32}}{Kx}+\frac{\Omega_{32}\epsilon_{23}}{7}\right)+N_{4}\left(\frac{A_{42}}{Kx}+\frac{\Omega_{42} \epsilon_{24}}{9}\right)\nonumber\\
	  +N_{5}\left(\frac{A_{52}}{Kx}+\frac{\Omega_{52} \epsilon_{25}}{11}\right)&=&0\nonumber\\
\end{eqnarray}
\begin{eqnarray}
		N_{1}\frac{\Omega_{13}\epsilon_{31}}{3}+N_{2}\frac{\Omega_{23}\epsilon_{32}}{5}\nonumber\\
	-N_{3}\left(\frac{A_{31}+A_{32}}{Kx}+\frac{\Omega_{13}+\Omega_{23}+\Omega_{43}+\Omega_{53}}{7}\right)\nonumber\\
	+N_{4}\left(\frac{A_{43}}{Kx}+\frac{\Omega_{43}\epsilon_{34}}{9}\right)+N_{5}\left(\frac{A_{53}}{Kx}+\frac{\Omega_{53} \epsilon_{35}}{11}\right)&=&0\nonumber\\
\end{eqnarray}
\begin{eqnarray}
		N_{1}\frac{\Omega_{14}\epsilon_{41}}{3}+N_{2}\frac{\Omega_{24}\epsilon_{42}}{5}+N_{3}\frac{\Omega_{34}\epsilon_{43}}{7}\nonumber\\
	-N_{4}\left(\frac{A_{41}+A_{42}+A_{43}}{Kx}\right)\nonumber\\
	-N_{4}\left(\frac{\Omega_{14}+\Omega_{24}+\Omega_{34}+\Omega_{54}}{9}\right)\nonumber\\
		+N_{5}\left(\frac{A_{54}}{Kx}+\frac{\Omega_{54} \epsilon_{45}}{11}\right)&=&0\nonumber\\
\end{eqnarray}
\begin{eqnarray}
N_{1}\frac{\Omega_{15}\epsilon_{51}}{3}+N_{2}\frac{\Omega_{25}\epsilon_{52}}{5}+N_{3}\frac{\Omega_{35}\epsilon_{53}}{7}+N_{4}\frac{\Omega_{45}\epsilon_{54}}{9}\nonumber\\
		-N_{5}\left(\frac{A_{51}+A_{52}+A_{53}+A_{54}}{Kx}\right)\nonumber\\
	-N_{5}\left(\frac{\Omega_{15}+\Omega_{25}+\Omega_{35}+\Omega_{45}}{11}\right)&=&0\nonumber\\
\end{eqnarray}
The elements $\alpha_{ij}$ of $\tilde{A}$ are then:
\begin{eqnarray}
	    \alpha_{11}&=&\alpha_{12}=\alpha_{13}=\alpha_{14}=\alpha_{15}=1\nonumber
\end{eqnarray}
\begin{eqnarray}
	    \alpha_{21}&=&\frac{\Omega_{12}\epsilon_{21}}{3}\nonumber\\
	    \alpha_{22}&=&-\frac{A_{21}}{Kx}-\frac{\Omega_{12}+\Omega_{32}+\Omega_{42}+\Omega_{52}}{5}\nonumber\\
	    \alpha_{23}&=&\frac{A_{32}}{Kx}+\frac{\Omega_{32} \epsilon_{23}}{7}\nonumber\\
	    \alpha_{24}&=&\frac{A_{42}}{Kx}+\frac{\Omega_{42} \epsilon_{24}}{9}\nonumber\\
	    \alpha_{25}&=&\frac{A_{52}}{Kx}+\frac{\Omega_{52} \epsilon_{25}}{11}\nonumber
\end{eqnarray}
\begin{eqnarray}
	    \alpha_{31}&=&\frac{\Omega_{13}\epsilon_{31}}{3}\nonumber\\
			\alpha_{32}&=&\frac{\Omega_{23}\epsilon_{32}}{5}\nonumber\\
	    \alpha_{33}&=&-\frac{\left(A_{31}+A_{32}\right)}{Kx}+\nonumber\\
	     & & -\frac{\Omega_{13}+\Omega_{23}+\Omega_{43}+\Omega_{53}}{7}\nonumber\\
	    \alpha_{34}&=&\frac{A_{43}}{Kx}+\frac{\Omega_{43} \epsilon_{34}}{9}\nonumber\\
	    \alpha_{35}&=&\frac{A_{53}}{Kx}+\frac{\Omega_{53} \epsilon_{35}}{11}\nonumber
\end{eqnarray}
\begin{eqnarray}
	    \alpha_{41}&=&\frac{\Omega_{14}\epsilon_{41}}{3}\nonumber\\
			\alpha_{42}&=&\frac{\Omega_{24}\epsilon_{42}}{5}\nonumber\\
	 		\alpha_{43}&=&\frac{\Omega_{34}\epsilon_{43}}{7}\nonumber\\
	    \alpha_{44}&=&-\frac{\left(A_{41}+A_{42}+A_{43}\right)}{Kx}+\nonumber\\
      &	& -\frac{\Omega_{14}+\Omega_{24}+\Omega_{34}+\Omega_{54}}{9}\nonumber\\
	    \alpha_{45}&=&\frac{A_{54}}{Kx}+\frac{\Omega_{54} \epsilon_{45}}{11}\nonumber
\end{eqnarray}
\begin{eqnarray}
	    \alpha_{51}&=&\frac{\Omega_{15}\epsilon_{51}}{3}\nonumber\\
			\alpha_{52}&=&\frac{\Omega_{25}\epsilon_{52}}{5}\nonumber\\
	 		\alpha_{53}&=&\frac{\Omega_{35}\epsilon_{53}}{7}\nonumber\\
	 		\alpha_{54}&=&\frac{\Omega_{45}\epsilon_{54}}{9}\nonumber\\
		  \alpha_{55}&=&\frac{\left(A_{51}+A_{52}+A_{53}+A_{54}\right)}{Kx}+\nonumber\\
	    & & \frac{\Omega_{15}+\Omega_{25}+\Omega_{35}+\Omega_{45}}{11}
\end{eqnarray}
It is at this point that we can identify the function $f$ in equation 1. The population of the $j_{th}$ level $N_{j}$ is given by,

\begin{equation}
	N_{j}=(\tilde{A}^{-1})_{j1}N.
\end{equation}

\noindent
Using the notion of the matrix of cofactors $C_{ji}$ known as an adjugate matrix such that,

\begin{equation}
	  (A^{-1})_{ij}=\frac{C_{ji}}{\left|\tilde{A}\right|}
\end{equation}

\noindent
then the level populations can be written in the form,

\begin{equation}
	N_{j}=\frac{C_{j1}\Big(\Omega_{ij}(T_{e}),x(n_{e},T_{e})\Big)N}{\left|\tilde{A}\right|}
\end{equation}

\noindent
or alternatively, 

\begin{equation}
	N_{j}=f(n_{e},T_{e})
\end{equation}

\noindent
as implied by equation 1. We have deliberately not inserted the values of $A_{ji}$, $\Omega_{ji}$ and $\epsilon_{ij}$ up until now as they are estimated values that will evolve with the accuracy of quantum mechanical calculations currently being performed by the Iron and Opacity Projects. In its present form, the matrix equation $\tilde{A}\textbf{y}=\textbf{b}$ with coefficients $\alpha_{ij}$ is exact. This is the closed form for the set of simultaneous equations for the 5-level ion. We used the symbolic manipulation toolbox of MATLAB to find the inverse matrix $\tilde{A}^{-1}$ and the level populations. The full solution includes terms up to $(Kx)^{4}$ and is too big to reproduce here. In the online supplementary material for this paper, we provide a MATLAB m-file whose results can be included into astrophysical codes. In general, an $n$-level ion will contain terms up to $(Kx)^{n-1}$. The solution for the 3-level ion includes terms only up to $(Kx)^{2}$ and is much more compact. Furthermore, its exact solution is known (Seaton 1975). This then provides a vital algebraic check on our solution to the 5-level ion as we demonstrate in the next section.\\

\subsection{The 3-level ion}
Supressing all terms with indices equal to 4 and 5 reduces the general solution for the 5-level ion to that of the 3-level ion. In doing so, for ions of type $p^{2}$ such as [OIII] as shown in figure 1, this amounts to removing the fine splitting physics of the $^{3}P$ level. As a consequence, the excitation potentials $E_{2}$ and $E_{3}$ of the 3-level ion correspond identically with $E_{4}$ and $E_{5}$ of the 5-level ion while $E_{1}$ is taken to be ground.\\

\begin{figure}[h]
  \begin{center}
  \vspace{6cm}
  \includegraphics{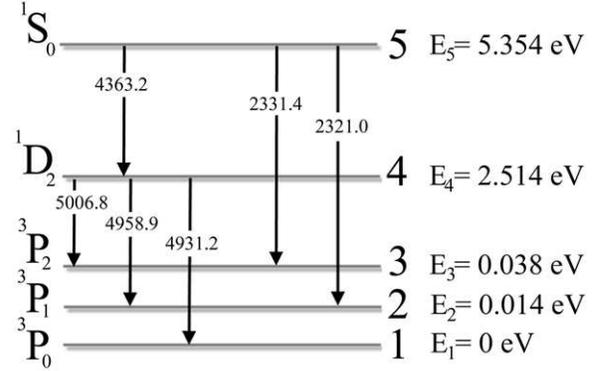}
  \caption{The 5-level [OIII] ion. Radiative transitions marked by arrows have their central emission wavelength in \AA}\protect\label{OIII}
  \end{center}
\end{figure}

\noindent
The exact level populations for the 3-level ion are then given by:
\begin{eqnarray}
N_{1}=\frac{3N}{D}\Big[35A_{21}A_{31}+35A_{21}A_{32}\nonumber\\
+Kx\Big(5A_{21}\Omega_{13}+5A_{21}\Omega_{23}+7A_{31}\Omega_{12}\nonumber\\
+7A_{31}\Omega_{32}+7A_{32}\Omega_{12}+7A_{32}\Omega_{32}-7A_{32}\Omega_{23}\epsilon_{32}\Big)\nonumber\\
+K^{2}x^{2}\Big(\Omega_{12}\Omega_{13}+\Omega_{12}\Omega_{23}+\Omega_{13}\Omega_{32}\nonumber\\
+\Omega_{23}\Omega_{32}-\Omega_{23}\epsilon_{32}\Omega_{32}\epsilon_{23}\Big)\Big]\\
\nonumber\newline\\
N_{2}=\frac{5N}{D}\Big[7Kx\Big(A_{31}\Omega_{12}\epsilon_{21}+A_{32}\Omega_{12}\epsilon_{21}+A_{32}\Omega_{13}\epsilon_{31}\Big)\nonumber\\
+K^{2}x^{2}\Big(\Omega_{12}\epsilon_{21}\Omega_{13}+\Omega_{12}\epsilon_{21}\Omega_{23}+\Omega_{13}\epsilon_{31}\Omega_{32}\epsilon_{23}\Big)\Big]\\
\nonumber\newline\\
N_{3}=\frac{7N}{D}\Big[Kx\Big(5A_{21}\Omega_{13}\epsilon_{31}\Big)\nonumber\\
+K^{2}x^{2}\Big(\Omega_{12}\epsilon_{21}\Omega_{23}\epsilon_{32}+\Omega_{12}\Omega_{13}\epsilon_{31}+\Omega_{13}\Omega_{31}\Omega_{32}\Big)\Big]
\end{eqnarray}
\newline
with the denominator $D$ term,
\newline
\begin{eqnarray}	
D=105A_{21}A_{31}+105A_{21}A_{32}\nonumber\\
+Kx\Big[15A_{21}\Omega_{13}+15A_{21}\Omega_{23}+21A_{31}\Omega_{12}+21A_{31}\Omega_{32}\nonumber\\
+21A_{32}\Omega_{12}+21A_{32}\Omega_{32}-21A_{32}\Omega_{23}\epsilon_{32}+35A_{21}\Omega_{13}\epsilon_{31}\nonumber\\
+35A_{31}\Omega_{12}\epsilon_{21}+35A_{32}\Omega_{12}\epsilon_{21}+35A_{32}\Omega_{13}\epsilon_{31}\Big]\nonumber\\
+K^{2}x^{2}\Big[3\Omega_{12}\Omega_{13}+3\Omega_{12}\Omega_{23}+3\Omega_{13}\Omega_{32}\nonumber\\
+3\Omega_{23}\Omega_{32}-3\Omega_{23}\epsilon_{32}\Omega_{32}\epsilon_{23}\nonumber\\
+5\Omega_{12}\epsilon_{21}\Omega_{13}+5\Omega_{12}\epsilon_{21}\Omega_{23}+5\Omega_{13}\epsilon_{31}\Omega_{32}\epsilon_{23}\nonumber\\
+7\Omega_{12}\Omega_{13}\epsilon_{31}+7\Omega_{12}\epsilon_{21}\Omega_{23}\epsilon_{32}+7\Omega_{13}\epsilon_{31}\Omega_{32}\Big].
\end{eqnarray}

\subsection{Line ratios of the 3-level ion}
The intensity of a radiatively-emitted line produced by collisional excitation of ions in a nebula can be written down directly from the level populations by noting that for any emission line, the emission rate $j(i,k)$ of line photons resulting from a downward transition $i\longrightarrow k$ is given by,
\begin{equation}
	4\pi j(i,k)=A_{ik}N(X^{l})N_{i}h\nu_{ik}
\end{equation}
where $X^{l}$ is the ion species e.g. $O^{2+}$ and $h\nu_{ik}$ corresponds to the energy difference $\Delta E$ between excitation potentials $E_{i}$ and $E_{k}$. The number of ions relative to ionised Hydrogen is given by,
\begin{equation}
\frac{N(X^{l})}{N(H^{+})}=\frac{I(i,k)}{I(H_{\beta})}\frac{j(H_{\beta})}{j(i,k)}.
\end{equation}
As stated in the introduction, there are standard methods that use optical emission line intensity ratios to measure electron temperature (Menzel et al. 1941) and electron density (Aller et al. 1949; Seaton 1954), including the simultaneous determination of both temperature and density (Dinerstein, Lester \& Werner 1985). Comparisons have also been made between various density indicators (Copetti \& Writzl 2002) and temperature indicators (Taylor \& V\'ilchez 2007d) as well as methodologies for the comparison between theory and observation (Seaton 1960; Taylor \& D\'iaz 2007a). In particular, the line ratio $R_{23}$,
\begin{equation}
	R_{23}\equiv\frac{I(\lambda4959)+I(\lambda5007)}{I(\lambda4363)},
\end{equation}
is a well known temperature diagnostic in ionised gaseous nebulae (Osterbrock \& Ferland 2006) and is based on the ion [OIII] (Pagel et al. 1979) whose first 5 levels above ground are shown in figure 1. In taking the ratio of emission lines, the number densities of the ion and ionised Hydrogen cancel leaving an expression in terms of atomic and ionic constants only. So, for example, for the 3-level ion, $R_{23}$ is given by,
\begin{eqnarray}
	R_{23}&=&\frac{j(2,1)_{5007}+j(2,1)_{4959}}{j(3,2)_{4363}}\nonumber\\
		    &=&\frac{N_{2}\Delta E_{21}}{N_{3}\Delta E_{32}}\Big[\frac{A_{21}(5007)+A_{21}(4959)}{A_{32}(4363)}\Big]
\end{eqnarray}
Many such line ratios of the ions of He, C, N, O, Ar, S, Ni and Cl are used as diagnostics of the physical conditions in photo-ionised gaseous nebulae (application of the formulae derived here for the n-level ion to a large collection of such diagnostics is in preparation(Taylor \& V\'ilchez 2007b).\\
\newline\noindent
For the purposes of the present discussion, we consider the ratio of the level populations $N_{3}$ and $N_{2}$ (although nowadays the inverse $N_{2}/N_{3}$ for the ratio of nebular to auroral lines is customary) so as to compare our exact solution for the 3-level ion with the  solution obtained by Seaton(1975):
\begin{eqnarray}
\frac{N_{3}}{N_{2}}=
\begin{array}{c}
\frac{7N}{D}\Big[Kx\Big(5A_{21}\Omega_{13}\epsilon_{31}\Big)\nonumber\\
+K^{2}x^{2}\Big(\Omega_{12}\epsilon_{21}\Omega_{23}\epsilon_{32}+\Omega_{12}\Omega_{13}\epsilon_{31}+\Omega_{13}\Omega_{31}\Omega_{32}\Big)\Big]\nonumber\\
\\
\hline\\
\frac{5N}{D}\Big[7Kx\Big(A_{31}\Omega_{12}\epsilon_{21}+A_{32}\Omega_{12}\epsilon_{21}+A_{32}\Omega_{13}\epsilon_{31}\Big)\nonumber\\
+K^{2}x^{2}\Big(\Omega_{12}\epsilon_{21}\Omega_{13}+\Omega_{12}\epsilon_{21}\Omega_{23}+\Omega_{13}\epsilon_{31}\Omega_{32}\epsilon_{23}\Big)\Big]\\
\end{array}
\end{eqnarray}
Extracting the term $\left(\Omega_{13}\epsilon_{31}\right)$ from the numerator and
$\left(\Omega_{12}\epsilon_{21}\right)$ from the denominator and calcelling like terms we get the result,
\begin{equation}
	\frac{N_{3}}{N_{2}}=
	\frac{\Omega_{13}}{\Omega_{12}}\frac{\epsilon_{31}}{\epsilon_{21}}
	\frac{A_{21}}{A_{32}}C\frac{\left[1+\frac{Kx}{A_{21}}\Psi_{3}\right]}{\left[1+\frac{Kx}{A_{32}}\Psi_{2}\right]}
\end{equation}
where,
\begin{eqnarray}
	C&=&\frac{1}{1+\frac{A_{31}}{A_{32}}+\frac{\Omega_{13}\epsilon_{31}}{\Omega_{12}\epsilon_{21}}}\nonumber\\
\Psi_{2}&=&C\left(\frac{\Omega_{12}\epsilon_{21}\Omega_{13}+\Omega_{12}\epsilon_{21}\Omega_{23}+\Omega_{13}\epsilon_{31}\Omega_{32}\epsilon_{23}}{7\Omega_{12}\epsilon_{21}}\right)\nonumber\\
\Psi_{3}&=&\frac{\Omega_{12}\epsilon_{21}\Omega_{23}\epsilon_{32}+\Omega_{12}\Omega_{13}\epsilon_{31}+\Omega_{13}\epsilon_{31}\Omega_{32}\epsilon_{23}}{5\Omega_{13}\epsilon_{31}}.
\end{eqnarray}
This is the same result as that obtained for the exact 3-level ion (Seaton 1975: equation 1.11).\\

\section{Theoretical, Observational and Experimental checks}
As with all new theoretical results, it is vital to perform checks. In the previous section we have verified algebraically that the equations for the 5-level ion (calculated from the full n-level ion equations), when reduced to the case of the 3-level ion, reproduce earlier results. In this section, we perform two additional checks. By considering the dimensions of the physical variables, we perform a check on the dimensional homogeneity of the solutions. Then, by inserting the most up-to-date quantum-mechanically calculated values for $A_{ji}$, $\epsilon_{ij}$ and $\Omega_{ji}$, we compare the values of the forbidden line ratio $R_{23}$ from the exact equations for the 5-level ion with those obtained from: a) observations of the benchmark planetary nebulae Abell 39, b) 3D Monte-Carlo photoionisation modelling of A39, c) numerical approximations provided by the nebular diagnostic software TEMDEN (De Roberties et al. 1987), and d) the exact 3-level ion.\\

\subsection{Dimensional consistency of the equations}
There are 7 units in the SI: mass $M[Kg]$, length $L[m]$, time $T[s]$, temperature $\theta[K]$, electrical current $I[A]$, concentration $N[mol]$ and light intensity $J[cd]$. In all equations of theoretical and empirical science, dimensional homogeneity for any set of physical variables $\{x_{i}\}$ must satisfy, 
\begin{equation}
	[x_{1}^{\epsilon_{1}}x_{2}^{\epsilon_{2}}...x_{n}^{\epsilon_{n}}]=1.
\label{homogeneity}
\end{equation}
for causality to be satisfied (Taylor et al. 2007c). The dimensions of the various physical quantities and parameters presented in our derivation are,
\begin{eqnarray}
\left[n_{e}\right]=\left[N_{i}\right]&=&L^{-3}\nonumber\\
\left[q_{ji}\right]=\left[q_{ij}\right]&=&L^{3}T^{-1}\nonumber\\
\left[n_{e}N_{i}q_{ji}\right]=&=&L^{-3}T^{-1}\nonumber\\
\left[A_{ji}\right]&=&T^{-1}\nonumber\\
\left[x(n_{e},T_{e})\right]&=&L^{-3}\Theta^{-1/2}\nonumber\\
\left[K\right]&=&L^{3}T^{-1}\Theta^{1/2}\nonumber\\
\left[\epsilon_{ji}\right]=\left[\omega_{j}\right]=\left[\Omega_{ji}\right]=\left[\alpha_{ij}\right]&=&1\nonumber
\end{eqnarray}
such that,
\begin{eqnarray}
\left[\frac{A_{ji}}{Kx}\right]=\left[\frac{Kx}{A_{ji}}\right]=\left[C\right]=\left[\Psi_{2}\right]=\left[\Psi_{3}\right]&=&1\nonumber
\end{eqnarray}
All equations were thoroughly checked for dimensional homogeneity with these dimensional relations and we did not find any inconsistencies.\\

\subsection{The forbidden line ratio $R_{23}$}
We have used observations of the spherically-symmetrical planetary nebula Abell 39 to make comparisons with our theoretical predictions. Although discovered by George Abell in his 1957 survey of the Southern Hemisphere (Abell 1966), A39 has only recently been subject to thorough spectrophotometry(Jacoby, Ferland \& Korista 2000) and 3D photoionisation modelling(Taylor et al. 2007e). From the detailed measurements made by George Jacoby and co-workers at Kitt Peak, the redenning-corrected line ratio $R_{23}$ averaged over the whole nebula, given by equation 28 equals,
\begin{equation}
		R_{23}=\frac{3.98\pm0.182+11.31\pm0.524}{0.24\pm0.019}=64.24\pm6.58
\end{equation}
with all fluxes $I(\lambda)$ relative to $I(H_{\beta})$. Using the 3D Monte-Carlo photoionisation code MOCASSIN (Ercolano et al. 2003) and a best-fit model to the observed spectrum(Taylor et al. 2007e), the following value, well within the observational error, was obtained,
\begin{equation}
	R_{23}=\frac{3.51+10.47}{0.21}=66.57.
\end{equation}

\begin{table*}
  \caption{Observed, simulated and theoretical values of the [OIII] line ratio $R_{23}$}\label{R23}
  \begin{tabular}{@{\vrule height 10.5pt depth4pt  width0pt}l|l|l|lllll}
  \hline
                            &      	        &				        &                &          & $R_{23}$ &         &\\
  \cline{4-8}
  CASE STUDY                & $n_{e}$       & $T_{e}$       & Observed       & MOCASSIN & TEMDEN   & 5-level & 3-level\\
  \hline
  Spectrophotometry of A39  & $n_{e}=30$    & $T_{e}\approx15400$ & $64.24\pm6.58$ &          & $65.16$  & $65.47$ & $64.92$\\
  (Jacoby et al. 2000)&&&&&&\\
  &&&&&&\\
  3D MOCASSIN models of A39 &               &               &                &          &          &				 & \\
  (Taylor et al. 2007e)&&&&&&\\

  (3-shell density profile) & $\left\langle n_{e}\right\rangle=11.94$ & $\left\langle T_{e}\right\rangle=15896$ &                & $66.57$  &          $60.93$  &	$61.98$ & $61.02$\\
  (uniform density)         & $n_{e}=30$    & $\left\langle T_{e}\right\rangle=14997$ &                & $69.91$  & $69.05$  & $70.25$ & $69.22$ \\
  &&&&&&\\
  QM Calculations of $\Omega_{ji}$  &               &               &                &          &          &				 & \\
  (Lennon \& Burke 1994)&&&&&&\\
  $log(T_{e})=4.0$ 					& $n_{e}=30$    & $T_{e}=10000$ &                &          & $209.77$ & $213.40$ & $203.11 $ \\
  $log(T_{e})=4.2$          & $n_{e}=30$    & $T_{e}=15848$ &                &          & $61.27$  & $62.37$  & $60.34 $ \\
  \end{tabular}
\end{table*}

In Table 1, we calculate the value of $R_{23}$ for the 5-level ion,

\begin{eqnarray}
	R_{23}=\frac{N_{4}}{N_{5}}\Big[\frac{\Delta E_{43}A_{43}+\Delta E_{42}A_{42}}{\Delta E_{54}A_{54}}\Big]
\end{eqnarray}
\noindent
using our exact theory, inserting the most up-to-date values of the atomic and ionic constants $A_{ij}$, $\Omega_{i,j}$ and $\epsilon_{ji}$ (Osterbrock \& Ferland 2006 and references therein), for representative values of electron density and temperature deduced empirically from the observations and simulations of A39. In addition, we provide results at a lower electron temperature of 10,000$K$ as a more extreme test case. The "theoretical" value of $R_{23}$ (calculated from the exact theory) is compared with that obtained from a) observations, b) 3D simulations using MOCASSIN, c) the first order 5-level ion of TEMDEN and d) the exact 3-level ion.\\
\newline\noindent
At electron densities of $n_{e}=30cm^{-3}$ all of the results are within the observational standard error apart from the low temperature case ($T_{e}=10000$) where, even here, there is consistency between the exact 5-level theory, TEMDEN and the exact 3-level theory.  In the case of the best-fit 3D model with MOCASSIN and using a 3-shell density profile having an average electron density of $<n_{e}>=11.94cm^{-3}$, the results are more dispersed with the 3D code producing a value of $R_{23}$ closer to the observations. As we have mentioned, this may be due to uncertainties in the values of the atomic constants $A_{ij}$, $\Omega_{ij}$ and $\epsilon_{ji}$. What is interesting is that the value of $R_{23}$ from the exact 5-level ion treatment is consistently higher than both the exact 3-level ion and also the first order approximation to the 5-level ion of TEMDEN. The effect of including all of the atomic physics provided by the exact theory does not appear to play a very significant role at these lower excitation stages nor at moderate values of plasma density and temperature.

\section{Conclusions}
We have shown that it is fairly straight-forward to generalise the calculation of ion level populations from the formal general statistical equilibrium equations. The matrix formulation presented here together with the algebraic power offered by symbolic manipulation software means that it has been possible to obtain an exact solution for the n-level ion. As a check on this result we have shown that, when indices for levels 4 and 5 for the 5-level ion are supressed, the exact solution for the 3-level ion is recovered. Iterations to higher level ions are straight-forward and we hope that these results will make the incorporation of more ionic transitions into existing astrophysics software easier for the study of very high ionisation objects such as AGN and SN. The 3-fold comparison between observations, theory and simulations appear to be consistent. In Paper II we will investigate a set of line ratio diagnostics for low, medium and high zones of ionisation based on the exact solution obtained here.

\section*{Acknowledgments}
MT would like to thank Luis Colina Robledo for making available a copy of Aller's 1984 book, the "Physics of Thermal Gaseous Nebulae" which provided the initial inspiration for the calculations performed here, Jos\'e Cernicharo for kindly facilitating this work and Angeles D\'iaz and the members of the Grupo de Astrof\'isica of the Universidad Aut\'onoma de Madrid for their hospitality and support. This work was partly funded by the project "Estallidos de Formación Estelar en galaxias" (AYA2001-3939-C03) from the Spanish ministry of science and is dedicated to the memory of Donald E. Osterbrock, Bernard E.J. Pagel and Michael J. Seaton who have done so much to advance this important field.


\bibliography{alpha}

\begin{thebibliography}{99}


\bibitem{ABELL1966}{Abell 1966} Abell G.O., 1966, ApJ, 144, 259
\bibitem{ALLER1949} Aller L.H., Ufford C.W., Van Vleck J.H., 1949, ApJ, 109, 42
\bibitem{ALLER1984} Aller L.H., ,1984, Physics of thermal gaseous nebulae. Reidel Press, Holland
\bibitem{COPETTI2002} Copetti M.V.F., Writzl B.C., 2002, A\&A, 382, 282
\bibitem{DEROBERTIS1987} De Robertis M.M., Dufour R.J., Hunt R.W., 1987, JRASC, 81, 195
\bibitem{DINERSTEIN1985} Dinerstein H.L., Lester D.F., Werner M.W., 1985, ApJ, 291, 561
\bibitem{ERCOLANO2003} Ercolano B., Barlow M.J., Storey P.J., Liu X.W., 2003, MNRAS, 340, 1136
\bibitem{HEBB1940} Hebb M.H., Menzel D.H., 1940, ApJ, 92, 408
\bibitem{JACOBY2000}	Jacoby G., Ferland G.J., Korista K.T., 2000, AAS, 197, 616
\bibitem{KAFATOS1980} Kafatos M., Lynch J.P., 1980, ApJS, 42, 611
\bibitem{LENNON1994} Lennon D.J., Burke V.M., 1994, AASS, 103, 273
\bibitem{MENDOZA1983} Mendoza C., 1983, in IAU Symp. 103, Planetary Nebulae. 83
\bibitem{MENZEL1941} Menzel D.H., Aller L.H., Hebb M.H., 1941, ApJ, 93, 230
\bibitem{MENZEL1962} Menzel D.H., ed., 1962, Physical processes in ionised plasmas. Dover Press, New York
\bibitem{OSTERBROCK1967} Osterbrock D.E., Seaton M.J., 1967, ApJ, 125, 66
\bibitem{OSTERBROCK2006} Osterbrock D.E., Ferland G.J., 2006, Astrophysics of gaseous nebulae and active galactic nuclei. University Science Books, USA
\bibitem{PAGEL1979} Pagel B.E.J., Edmunds M.G., Blackwell D.E., Chun M.S., Smith G., 1979, MNRAS, 189, 95
\bibitem{PMD2005}P\'erez-Montero E., D\'iaz A.I., 2005, MNRAS, 361, 1063
\bibitem{SEATON1954} Seaton M.J., 1954, Ann d'Ap, 17, 74
\bibitem{SEATON1960} Seaton M.J., 1960, Rep. Prog. Phys., 23, 313
\bibitem{SEATON1968} Seaton M.J., 1968, Adv. Atom. Mol. Phys., 4, 331
\bibitem{SEATON1975} Seaton M.J., 1975, MNRAS, 170, 475
\bibitem{SHAW1995} Shaw R.A., Dufour R.J., 1995, PASP, 107, 896
\bibitem{SPITZER1948} Spitzer L., 1948, ApJ, 107, 6
\bibitem{TAYLOR2007a} Taylor M., D\'iaz A.I., 2007, PASP, in press
\bibitem{TAYLOR2007b} Taylor M., V\'ilchez J.M., 2007, MNRAS, in preparation
\bibitem{TAYLOR2007c} Taylor M., D\'iaz A.I., J\'odar-S\'anchez L.A., Mic\'o R.J., 2007, PNAS, submitted preprint (arXiv:0709.3584)
\bibitem{TAYLOR2007d} Taylor M., V\'ilchez J.M., 2007, MNRAS, in preparation
\bibitem{TAYLOR2007e} Taylor M., Jacoby G., Rauch T., Ercolano B., D\'iaz A.I., 2007, MNRAS, in preparation

\end{thebibliography}

\label{lastpage}
\end{document}